\documentclass[prc,superscriptaddress,nofootinbib,onecolumn,12pt]{revtex4-2}

\usepackage{graphicx}
\usepackage{amssymb}
\usepackage{amsmath}
\usepackage{dcolumn}
\usepackage{bm}
\usepackage{multirow}
\usepackage{color}
\usepackage[normalem]{ulem}
\usepackage[draft]{changes} 
\usepackage{makecell}

\newcommand{\cev}[1]{\reflectbox{\ensuremath{\vec{\reflectbox{\ensuremath{#1}}}}}}
\def\e{{\mathrm{e}}}

\def\vk{{\bm k}}
\def\vp{{\bm p}}

\def\be{\begin{eqnarray}}
\def\ee{\end{eqnarray}}

\begin{document}

\title{Elastic 
p-$^{12}$C scattering by using a cluster effective field theory}
\author{Eun Jin In}
\affiliation{Department of Energy Science, Sungkyunkwan University, Suwon 16419, Korea}
\affiliation{Center for Exotic Nuclei Studies, Institute for Basic Science, Daejeon 34126, Korea}
\affiliation{Lawrence Livermore National Laboratory, Livermore, CA 94551, USA}
\author{Tae-Sun Park}
\email{tspark@ibs.re.kr}
\affiliation{Center for Exotic Nuclei Studies, Institute for Basic Science, Daejeon 34126, Korea}
\author{Young-Ho Song}
\affiliation{Institute for Rare Isotope Science, Institute for Basic Science, Daejeon 34000, Korea}
\author{Seung-Woo Hong}
\affiliation{Institute for Rare Isotope Science, Institute for Basic Science, Daejeon 34000, Korea}
\affiliation{Department of Physics, Sungkyunkwan University, Suwon 16419, Korea}

%

\date{\today}

\begin{abstract}
The elastic p-$^{12}$C scattering at low energies is studied 
by using a cluster effective field theory (EFT),
where
the low-lying resonance states ($s_{1/2}$, $p_{3/2}$, 
$d_{5/2}$) of 
$^{13}$N
are treated as pertinent degrees of freedom.
The low-energy constants of the Lagrangian are expressed in terms of the Coulomb-modified effective range parameters, 
which are 
determined 
to reproduce
the experimental data
for the differential cross-sections.
The resulting 
theoretical predictions 
agree very well with the experimental data.
The resulting theory
is shown to give us 
almost identical
phase shifts
as obtained from
the $R$-matrix approach.
The role of the ground state of ${}^{13}$N below the threshold
and the 
next-to-leading order in the EFT power counting are
also discussed.
\end{abstract}

\maketitle

\section{INTRODUCTION}

%
%
The radiative proton capture reaction of carbon-12, 
$^{12}$C(p,$\gamma$)$^{13}$N, 
plays an important role 
in the CNO cycle~\cite{Ago2020}.
That is,
the chain of
$^{12}$C(p,$\gamma$)$^{13}$N($\beta^+$)$^{13}$C reactions 
increases
the $^{13}$C abundance
and 
hence
the $^{13}$C($\alpha$,n)$^{16}$O reaction 
that acts as a neutron source 
in the asymptotic giant branch (AGB) stars~\cite{mowlavi1998}.
However,
the reaction cross-section at astrophysical energies
is difficult to determine experimentally due to the Coulomb barrier.
Thus,
employing a theoretical model is 
useful
to extrapolate the cross-section
at very low astrophysical energies.

The reaction has been studied in diverse theoretical approaches,
which include
potential models like potential cluster model (PCM)~\cite{kabir2020}, 
single-particle model~\cite{huang2010}, 
distorted wave Born approximation (DWBA)~\cite{li2010}, 
and the phenomenological R-matrix theory~\cite{Azu2010}.

$R$-matrix theory provides a reliable theoretical tool to determine $S$ factors
at low energies.  
However,
the cluster effective field theory (EFT) 
can be an alternative approach to the $R$-matrix theory.
The cluster EFT provides a powerful framework to describe the low-energy system 
by exploiting the scale separation of the system.
The EFT uses
the systematic expansion scheme of the theories,
and
thus allows
improved calculations with well-defined error estimates.
The cluster EFT~\cite{ber2002} has been used for the analysis of
diverse nuclear systems, including
the one-neutron halo nucleus $^{19}$C~\cite{ach2013},
one-proton halo nuclei $^{17}$F and $^{8}$B~\cite{Ryb2014,zhang2015}.
It has also been applied to non-halo systems with the existence of scale separation 
such as the resonant $\alpha$-$\alpha$ scattering Ref.~\cite{higa2008}
and $^{12}$C-$\alpha$ scattering\cite{ando2016,ando2018}.

In the present work, 
we analyze the differential cross section for elastic p-$^{12}$C scattering
in the cluster EFT,
which 
is important for the EFT-description
of the $^{12}$C(p,$\gamma$)$^{13}$N reaction.
In addition, very accurate experimental data on elastic scattering exist, 
which 
is
useful to guide and test our theoretical approach.
%
As we will show later,
the reaction is 
dominated by
the three low-lying resonance states
of $^{13}$N
with $J_{\pi}=1/2^+,3/2^-$ and $5/2^+$,
which will be treated
as pertinent degrees of freedom
of our cluster EFT.
The ground state ($J_{\pi}=1/2^-$) of $^{13}$N
lies below the threshold energy,
and plays only a minor role,
as was also studied in the 
$R$-matrix analysis~\cite{Azu2010}.
We will quantify its importance by comparing
cluster EFTs with and without the ground state.

This paper is organized as follows:
In Section~\ref{Sec:Method}, 
the cluster EFT formalisms for $s$-, $p$- and $d$-wave interactions
of elastic p-$^{12}$C scattering
and renormalization conditions
are given.
In Section ~\ref{Sec:Results},
we present
the results of renormalization, and
phase shift analysis and comparison with 
the $R$-matrix are also discussed.
In Section~\ref{Sec:Conclusions},
we give a conclusion and
discuss a possible future work.

\section{Cluster EFT for $s$-, $p$- and $d$-wave interactions \label{Sec:Method}}

In this section, we present our formalism
for elastic p-$^{12}$C scattering in the framework of cluster EFT.
Many useful discussions of our formalism
can 
be found in
Refs.~\cite{ber2002,Hammer2020}.

\subsection{Scale separation and Lagrangian} 

\begin{figure} [htb]
\centering
\resizebox{\columnwidth}{!}{%
\includegraphics[]{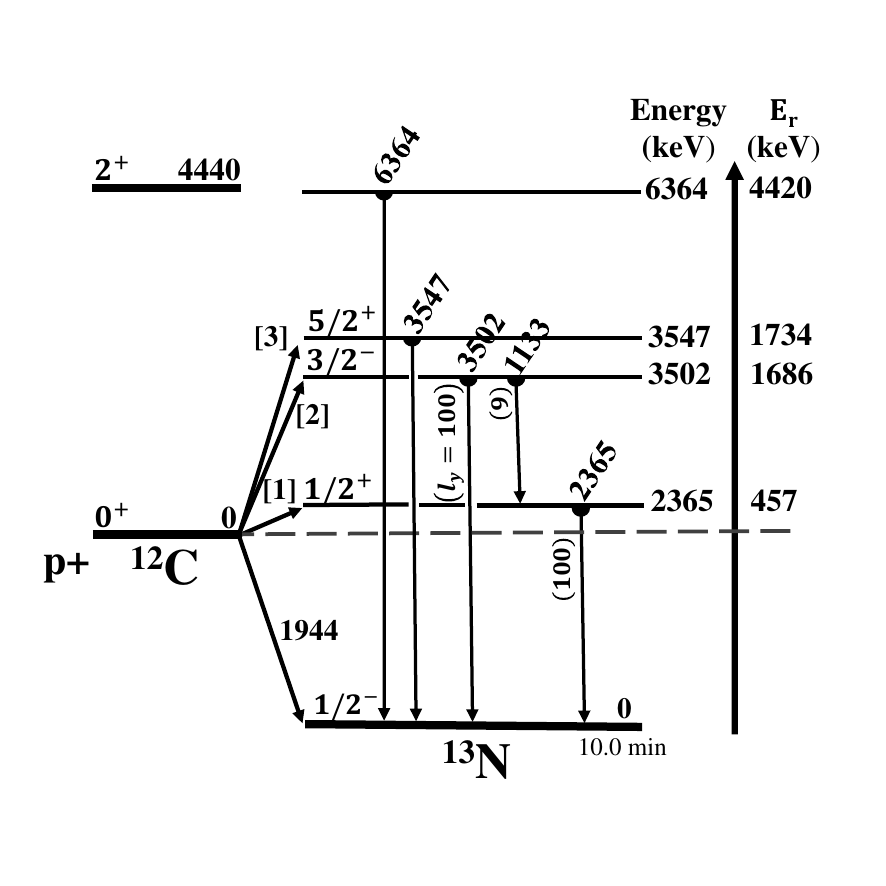}}%
\caption{\label{PSR}
Level scheme of $^{13}$N (Not to scale)}
\end{figure}

Figure~\ref{PSR} depicts the level scheme of 
the compound nucleus $^{13}$N.
The three low-lying resonance states 
with $J^{\pi}=1/2^{+},\,3/2^{-}$,
and $5/2^{+}$ of $^{13}$N
are taken as pertinent degrees of freedom of the theory.
Their respective excitation energies
are $E_r=0.457$, 1.686, 
and 1.734 MeV, 
with
corresponding momenta of 
$\sqrt{2m_\mathrm{R} E_r}\,=\,28,\,54,
\,55$ MeV,
where 
$m_\mathrm{R}$ is the reduced mass of the p-$^{12}$C system.
These momenta are characterized by the scale denoted as $k_{lo}$,
which is regarded as small compared with
the high momentum scale $k_{hi}$.
The natural choice for $k_{hi}$ is 
the momentum corresponding 
to the core excitation $\sqrt{2m_\mathrm{R} E^*}\,\sim\,90$ MeV,
where $E^*=4.44$ MeV
is the first excitation energy of $^{12}$C.
The EFT is then expanded with increasing power of
the ratio $k_{lo}/k_{hi}\,=\,(0.3 \sim 0.6)$.
The scale associated with the 
Coulomb interaction, 
$k_{\mathrm{C}}=Z_{\mathrm{C}} \alpha_{EM} m_\mathrm{R}\sim 38$ MeV,
is numerically comparable to $k_{lo}$,
where $Z_C=6$ and
$\alpha_{EM}\simeq 1/137.036$ is the fine structure constant.
The ground state of $^{13}$N with
$J^\pi= 1/2^{-}$ is 
a sub-threshold state
located below the threshold,
$E_r= -1.944$ MeV, and its role will be
discussed later.

The effective Lagrangian for the system can be written as~\cite{ryberg2014,bedaque2003,braun2019}
\be
\label{Lag}
\mathcal{L} &&=
\psi_p^{\dag} \left(iD_t + \frac{\vec{D}^2}{2m_p}\right) \psi_p
+ \psi_c^{\dag} \left(iD_t + \frac{\vec{D}^2}{2m_c}\right) \psi_c 
\nonumber \\
&&+ \sum_{x} d_{x}^{\dag}
\left[ \Delta_{x}+ \sum^{N_x}_{n=1} \nu_{n,x}
\left(iD_t + \frac{\vec{D}^{2}}{2m_{tot}} \right)^n \right] d_x
\nonumber \\
&& -\,
\sum_{x} g_{x}
\left[\, d_{x}^{\dag} 
\left[ \psi_p \,i\, \overleftrightarrow{\nabla} \, \psi_c \,\right]_x
+ \mathrm{h.c.} \right] + \cdots,
\ee
where $\psi_p$, $\psi_c$ and $d_x$
are the proton, $^{12}$C
and the dicluster field, respectively,
with the subscript $d_x$ denoting the 
total angular momentum and parity of the dicluster, $x=J^{\pi}$.
Their masses are denoted as $m_p$, $m_c$ and $m_{tot}=m_{p}+m_{c}$,
respectively,
and the covariant derivatives are defined as 
$D_{\mu}=\partial_{\mu}+ie\hat{Q}A_{\mu}$, 
where $\hat{Q}$ is the charge 
operator.
The parameters $\Delta_x$ and $g_x$
represent 
the residual masses and coupling constants 
of field $d_x$,
respectively. 
The index $n$ 
is 1 for $s$- and $p$- waves,
and runs up to 2 for $d$- wave.
The $\nu_{1,x}$ in the kinetic term of the dicluster field are chosen 
as $\pm 1$ to be a sign 
related to the effective range~\cite{ber2002},
while the $\nu_{2,x}$ in the 2nd-order kinetic term for $d$-wave
is needed for renormalization. 
At LO, we have therefore two low-energy constants (LECs) 
for $s$- and $p$-waves,
and three low-energy constants for $d$-wave.
As we will show 
shortly,
these LECs are to be related to the effective range parameters.

The projection of the operator
$\psi_p \,i \overleftrightarrow{\nabla} \psi_c = 
\psi_p \left(
m_c i \cev{\nabla} - m_p i \vec{\nabla}
\right) \psi_c/(m_p+m_c)$
to the $x=1/2^{+},\,3/2^{-},\,1/2^{-}$, and $5/2^{+}$ states
are given as~\cite{tanabashi2018}
%
\be
\left[\psi_p i\overleftrightarrow{\nabla} \psi_c \right]_{\frac{1}{2}^{+}}^m &&= 
\sum_{m_s} C_{0 0, \frac{1}{2} m_s}^{\frac{1}{2} m} \, \psi_p \psi_c, 
\nonumber \\
\left[\psi_p i\overleftrightarrow{\nabla} \psi_c \right]_{\frac{3}{2}^{-} \left(\frac{1}{2}^{-}\right)}^m &&= 
\sum_{\alpha, m_s} 
C_{1 \alpha, \frac{1}{2} m_s}^{\frac{3}{2} m \left(\frac{1}{2} m\right)} \, \psi_p i\overleftrightarrow{\nabla}_{\alpha} \psi_c, 
\nonumber\\
\left[\psi_p i\overleftrightarrow{\nabla} \psi_c \right]_{\frac{5}{2}^{+}}^m 
&&= 
\sum_{\alpha, \beta, m_l, m_s}
C_{2 m_l, \frac{1}{2} m_s}^{\frac{5}{2} m}  C_{1 \alpha, 1 \beta}^{2 m_l} 
 \psi_p 
 \nonumber\\
&& \times\
\frac12  \left(i\overleftrightarrow{\nabla}_{\alpha} i\overleftrightarrow{\nabla}_{\beta} 
+ i\overleftrightarrow{\nabla}_{\beta} i\overleftrightarrow{\nabla}_{\alpha}
\right)
\psi_c, 
\label{Cart}
\ee
where $m_s$ is the spin projection of the proton
and $C_{j_1 m_1, j_2 m_2}^{Jm}$ 
is a short notation for  
the Clebsch-Gordan coefficients
$\langle j_1 m_1, j_2 m_2| (j_1 j_2) jm\rangle$.
Here and hereafter, we use 
the Greek letters 
to denote spherical components that run
from $-1$ to $1$.
The conversion to Cartesian coordinates
for convenience in 
the calculation of the $d$-wave
can be found in Ref.~\cite{bedaque2003}.

\subsection{The irreducible self-energy and renormalization conditions \label{Sec:Cal}}

\begin{figure} [htb]
\centering
\resizebox{0.75\columnwidth}{!}{
\includegraphics[]{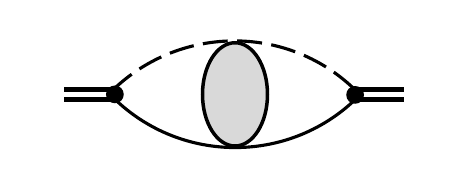}}%
\caption{\label{self}
Self-energy diagram of a dicluster.
The solid line denotes the core($^{12}$C) 
and the dashed line represents the proton field. 
The shaded bubble denotes the Coulomb Green's function.}
\end{figure}
%

The full dicluster propagator
of the dicluster $d_x$
reads
\be
iD_{x}(E)
=\frac{i}{\Delta_{x}+\sum_{n=1}^{N_x}\nu_{n,x}(E+i\epsilon)^n-\Sigma_{x}(E)},
\label{iDx}\ee
where $\Sigma_{x}(E)$ is the irreducible self-energy 
shown in  Fig.~\ref{self}.
The 
Coulomb interaction plays a crucial role at low-energy,
and is taken into account by the Coulomb Green's function.
Because each dicluster of $x$ in our consideration 
has a different orbital angular momentum $l$,
we will use $l$ and $x$ interchangeably hereafter.


%
\begin{figure} [htb]
\centering
\resizebox{0.55\columnwidth}{!}{
\includegraphics[]{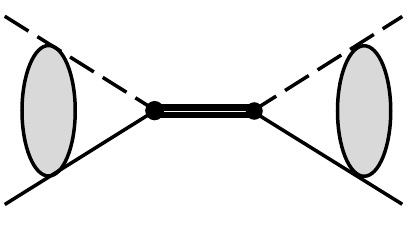}}%
\caption{\label{scattamp} 
Scattering amplitude for elastic p-$^{12}$C scattering. 
The notation is the same as in Fig.~\ref{self}}
\end{figure}

The elastic scattering amplitude 
$T_l$ for $s$-, $p$- and $d$-waves
are depicted in
Fig.~\ref{scattamp},
and
can be evaluated 
as~\cite{ando2018elastic} 
\be \label{tEFT}
T_l= g_l^2 D_l(E)e^{2i\sigma_l}k^{2l} \hat{C}_l^2(\eta),
\ee
where 
$\sigma_l=\mathrm{arg}\Gamma(l+1+i\eta)$,
and  
\be
\hat{C}_l(\eta) = \left| \Gamma(l+1+i\eta)\right| \e^{-\frac12 \pi\eta}/\Gamma(l+1),
\ee
which is 
the Gamow-Sommerfeld factor $C_l(\eta)$~\cite{abramowitz1984pocketbook,Ryb2014}
but normalized to unity when $\eta$ goes to zero.


The scattering amplitude in Eq.(\ref{tEFT}) 
can be matched with the effective range function
as~\cite{higa2008}
\be \label{tERE}
T_{l}(E)=- \frac{2\pi}{m_\mathrm{R}} 
\frac{k^{2l}e^{2i\sigma_l} \hat{C}_l ^2(\eta)}
{f_l(k)-2k_{\mathrm{C}}h_l(\eta)}.
\ee
Here, $f_l(k)$ 
is the Coulomb-modified 
effective range function (ERF)~\cite{berger,konig2013effective},
\be \label{kcot}
f_l(k) &\equiv&
k^{2l+1} \hat{C}_l(\eta)^2 \left(\mathrm{cot}\delta_l(k)- i \right)+2k_{\mathrm{C}}h_l(\eta) 
\\ \nonumber
&=&-\frac{1}{a_l}+\frac{1}{2}r_l k^2-\frac{1}{4} P_l k^4+\cdots,
\label{xl}\ee
where $\delta_l$ is the phase shifts relative to the regular Coulomb function
for angular momentum $l$,
$a_l$, $r_l$ and $P_l$ are the effective range parameters 
(scattering length, effective range and shape parameter),
and the function $h_l(\eta)$ is defined as~\cite{jackson1950}
\be
h_l(\eta)=k^{2l} \frac{\hat{C}_l(\eta)^2}{\hat{C}_0(\eta)^2} \left(\psi(i\eta)+\frac{1}{2i\eta}-\mathrm{log}(i\eta) \right),
\ee
where $\psi(z)=\Gamma'(z)/\Gamma(z)$ is 
the logarithmic derivative of the Gamma function.
%
Comparison of Eq.(\ref{tERE}) with Eq.(\ref{tEFT})
enables us to renormalize 
the LECs in terms of the effective range parameters.

\subsubsection{$S$-wave interaction}

The irreducible self-energy of $s$-wave dicluster can be expressed as
\be
\Sigma_0(E) &=& g_0^2 \int \frac{d^3 \vk d^3 \vk'}{(2\pi)^6} 
\left\langle\vk| G_{\mathrm{C}}(E)|{\vk'}\right\rangle
\nonumber\\
 &=& g_0^2 \int \frac{d^3 \vp}{(2\pi)^3} \frac{\psi_{\vp}(0) 
 \psi_{\vp}^*(0)}{E-{p^2}/2m_\mathrm{R} +i\epsilon},
\label{SigmaOE}
\ee
where $G_{\mathrm{C}}(E)$ is the Coulomb Green's function~\cite{kong2000coulomb},
%
\be
\left\langle{\bf r}|G_{\mathrm{C}}(E)|{\bf r'}\right\rangle
=\int \frac{d^3 \vp}{(2\pi)^3}\,
\frac{\psi_{\vp}({\bf r})\psi_{\vp}^*({\bf r'})}{E-\frac{{\vp}^2}{2m_\mathrm{R}}+ i\epsilon},
\label{GCr}\ee
and $\psi_{\vp}({\bf r})$ is the Coulomb wave function
\be
\psi_{\vp}({\bf r})=\sum_{l=0}^{\infty} (2l+1) i^l e^{i\sigma_l} 
\frac{F_l(\eta, p r)}{p r} P_l(\hat{p} \cdot \hat{r}),
\ee
$\eta=k_{\mathrm{C}}/p$, 
$F_l$
are
the regular 
Coulomb functions ~\cite{konig2013}.

%
%

%
%


The integral in Eq.(\ref{SigmaOE}) can be evaluated by using 
the power divergence subtraction (PDS) method~\cite{kong2000coulomb,kong1999proton,ando2007low},
\be
\Sigma_0(E)&=&
-g_0^2 \frac{k_C m_\mathrm{R}}{\pi}h_0(\eta)
+\Sigma_0^{\mathrm{div}},
\ee
and the 
divergent part $\Sigma_0^{\mathrm{div}}$ is energy-independent,
whose 
explicit form
can be found in Refs.~\cite{kong1999proton,ando2007low}.

The $s$-wave ERF with $l=0$
is then given as
\be\label{f_0k}
f_0(k)=
-\frac{2\pi}{g_0^2 m_\mathrm{R}}(\Delta_0+\Sigma_0^{\mathrm{div}})
-\frac{\pi \nu_0}{g_0^2 m_\mathrm{R}^2}k^2.
\ee
Comparison of Eq.(\ref{f_0k}) with
Eq.(\ref{tERE})
gives us
the renormalization conditions
\be
\frac{1}{a_0} &&= \frac{2\pi}{g_0^2m_\mathrm{R}} \left(\Delta_0+\Sigma^{\mathrm{div}} \right), 
\nonumber\\
r_0 &&= -\frac{2\pi\nu_0}{g_0^2 m_\mathrm{R}^2}.
\ee

\subsubsection{$P$-wave interaction}


By using a similar procedure as
for the
s-wave,
the irreducible self-energy of $p$-wave dicluster 
can be derived as~\cite{ber2002}, 
\be
\Sigma_1(E)&=&  \frac13 g_1^2 \int \frac{d^3\vp}{(2\pi)^3}
\frac{p^2 \hat{C}_1(\eta_p)^2}{E-\frac{p^2}{2m_\mathrm{R}}+i\epsilon}
\nonumber \\
 &=&  g_1^2 \frac{m_\mathrm{R}}{3\pi^2}
 \Big[
 -L_3-(k^2+k_\mathrm{C}^2)L_1 
 \nonumber \\
&& 
+k^2(k^2+k_\mathrm{C}^2) \int dp 
\frac{\hat{C}_0(\eta)^2}{k^2-p^2+i\epsilon} \Big]
\nonumber\\
&=&
\frac{g_1^2 m_\mathrm{R}}{6\pi} 
\left[-\frac{2}{\pi}L_3-\frac{2}{\pi}(k^2+k^2_\mathrm{C})L_1-2k_\mathrm{C} h_1(\eta) \right],
\ee
where 
\be
L_n &&=\int dp\,\hat{C}_0(\eta_p)^2 p^{n-1}.
\ee
It is then a simple task to show that
the resulting $p$-wave ERF reads
$f_1(k) = -\frac{1}{a_1} + \frac12 r_1 k^2$ with
%
\be
\frac{1}{a_1} &&=-\frac{6\pi}{m_\mathrm{R}}\left(\frac{\Delta_1}{g_1^2}-\frac{m_\mathrm{R}}{3\pi^2}L_3 \right.
\left. -\frac{m_\mathrm{R}}{3\pi^2}k_\mathrm{C}^2L_1 \right), 
\nonumber \\
r_1 &&= \frac{6\pi\nu_1}{g_1^2 m_\mathrm{R}^2}-\frac{4}{\pi}L_1.
\ee

\subsubsection{$D$-wave interaction}

%
By adopting the trick of using Cartesian representation 
of the $d$-wave vertex function~\cite{braun2019,brown2014}, 
the irreducible self-energy of $d$-wave dicluster 
can be evaluated as 
\be
\Sigma_2(E)&=&
g_2^2 \frac{2}{15} \int \frac{d^3p}{(2\pi)^3}
\frac{p^4 \hat{C}_2(\eta_p)^2}{E-\frac{p^2}{2m_\mathrm{R}}+i\epsilon}  
\nonumber\\
&=&\frac{g_2^2 m_\mathrm{R}}{15\pi} \left[\left(-\frac{8}{\pi}L_5-\frac{10}{\pi}k_\mathrm{C}^2L_3
-\frac{2}{\pi}k_\mathrm{C}^4L_1 \right) \right. 
\nonumber\\
& &\left. +\left(-\frac{8}{\pi}L_3-\frac{10}{\pi}k_\mathrm{C}^2L_1 \right)k^2
-\frac{8}{\pi}L_1k^4-2k_\mathrm{C} h_2(\eta) \right].
\ee
The corresponding
$d$-wave ERF is then given as
$f_2(k) = -\frac{1}{a_2} + \frac12 r_2 k^2 - \frac14 P_2 k^4$ with
\be
\frac{1}{a_2}&&=\frac{15\pi}{g_2^2 m_\mathrm{R}}\Delta_2 +\frac{8}{\pi}L_5
+\frac{10}{\pi}k_\mathrm{C}^2L_3
+\frac{2}{\pi}k_\mathrm{C}^4L_1, 
\nonumber \\
r_2 &&=-\frac{15\pi\nu_{1,2}}{g_2^2 m_\mathrm{R}^2}
-\frac{16}{\pi}L_3 -\frac{20}{\pi}k_\mathrm{C}^2L_1, 
\nonumber \\
P_2 &&= \frac{15\pi\nu_{2,2}}{m_\mathrm{R}^3g_2^2} 
+\frac{32}{\pi}L_1.
\ee

\section{Numerical results and discussion \label{Sec:Results}} 

\subsection{Fitting to experimental data \label{Sec:CompExp}}

In the previous section,
we have shown that the cluster EFT description with the LECs
is identical to the Coulomb-modified ERF
with a finite number of 
effective range parameters (ERPs),
and the remaining task is to determine the values of the parameters
from the experimental data.
We find that the fitting for the ERPs is 
complicated
due to the strong correlations 
between the
ERPs of the $p$ and $d$-waves,
which is caused 
mainly
by the fact that
their pole positions are very close to each other.
%
This problem can be avoided
by rewriting the effective range function 
as a series around the pole position,
\be \label{modERE}
f_l(k)
&&= -\frac{1}{a_l} + \frac{1}{2}r_l k^2 -\frac{1}{4} P_l k^4 + Q_l k^6 + \cdots \nonumber \\ 
 &&=\frac{1}{2}r_l^{'}(k^2-k_r^2) -\frac{1}{4}P_l^{'}(k^2-k_r^2)^2  \nonumber \\
 && \,\,\,\,+ \,Q_l^{'}(k^2-k_r^2)^3 + \cdots,
\ee
where $k_r^2\equiv m_\mathrm{R} E_r$,
$E_r$ being the resonance excitation energy,
and 
$(E_r, r_l^{'}, P_l^{'}, Q_l^{'}, \cdots)$ are 
another representation of the ERPs $(a_l, r_l, P_l, Q_l,\cdots)$.

The parameters $(E_r^{'}=E_r,\, r_l^{'},\, P_l^{'},\, Q_l^{'},\, \cdots)$ 
are determined by minimizing $\chi_\Lambda^2$ defined as 
\be
\chi_\Lambda^2=\sum_i^N 
\frac{\left|y_{i,\text{th}}-y_{i,\text{exp}}\right|^2}{
\Delta y_{i,\text{exp}}^2 + \Delta y_{i,\text{th}}^2
} 
\label{chi2}\ee
where $y_{i,\text{exp}}$ ($y_{i,\text{th}}$) is the experimental (theoretical)
differential cross sections at a given angle, 
$\Delta y_{i,\text{exp}}$ are error bars of the data.
Some of the data have very small $\Delta y_{i,exp}$,
and the usual chi-square is dominated by them.
As a regulator that takes into account the theoretical uncertainty,
we introduce
\be
\Delta y_{i,\text{th}}\equiv y_{i,\text{exp}}
\times \frac{k_i}{\Lambda},
\label{Lambdai}
\ee
where $\Lambda$ is a parameter.
While constructed in an ad-hoc manner,
this form is motivated by the fact that
the EFT description is 
less accurate at high momentum.
$\Delta y_{i,\text{th}}$
should not be
bigger than
the uncertainty of the theory,
and thus we
choose  $\Lambda=1$ GeV.
We find that 
the resulting parameters are found to be stable and
insensitive to the values of $\Lambda$,
while the value of $\chi_\Lambda^2$
increases with $\Lambda$.
%

So far, we have 
considered only
the leading order (LO) terms,
and the resulting theory turns out to be identical
to the Coulomb-modified effective range expansion
with the parameters
$(E_r, r_l^{'})$ for $s$- and $p$-waves 
and $(E_r, r_l^{'}, P_l^{'})$ for $d$-wave.
While we do not describe explicitly here,
going to the next order (or NLO)
with including one higher-order terms in the Lagrangian
is also identical to the 
effective range expansion
with one more term, that is,
%
$(E_r, r_l^{'}, P_l^{'})$ for $s$- and $p$-waves 
and $(E_r, r_l^{'}, P_l^{'}, Q_l^{'})$ for $d$-wave,
which we denote as
NLO.
We also consider the leading order calculation
where the $J^\pi=\frac12^-$ ground
state of ${}^{13}$N is also taken as
a pertinent degrees of freedom,
which we denote as LO+gs.
We thus have three sets of parameters,
LO, NLO and LO+gs.

\begin{figure} [htb]
\centering
\resizebox{\columnwidth}{!}{%
\includegraphics[]{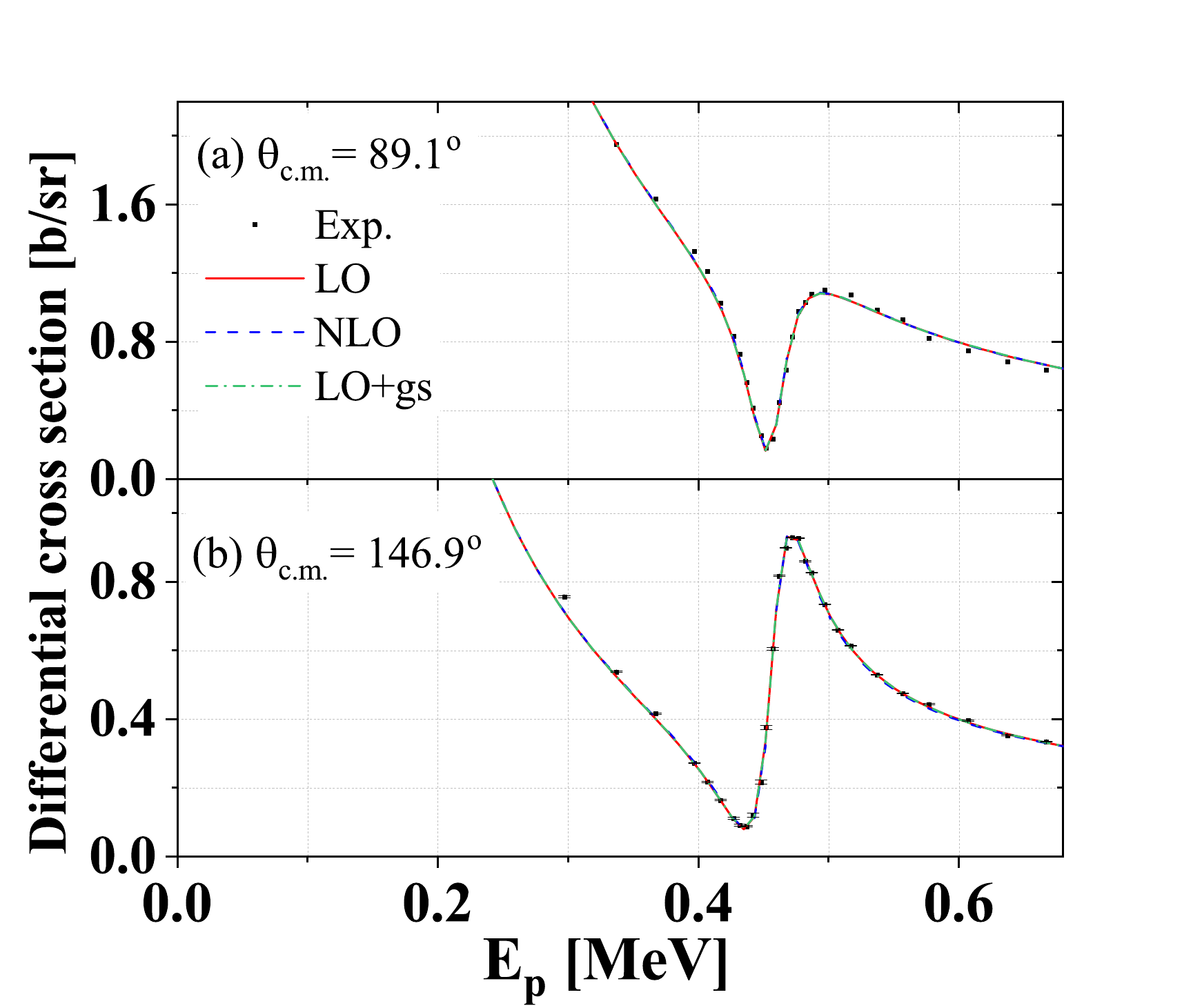}}%
\caption{\label{p12C_CSlow}
The differential cross section for elastic p-$^{12}$C scattering 
as a function of the proton energy 
$E_{p} < 0.7$ MeV
at two angles
(a) 89.1$^{\circ}$ and (b) 146.9$^{\circ}$. 
The red, blue, and green solid lines represent the EFT results 
at LO, NLO, and LO+gs.
The black circles represent the experimental data~~\cite{meyer1976}.}
\end{figure}
\begin{figure} [h]
\centering
\resizebox{0.98\columnwidth}{!}{%
\includegraphics[]{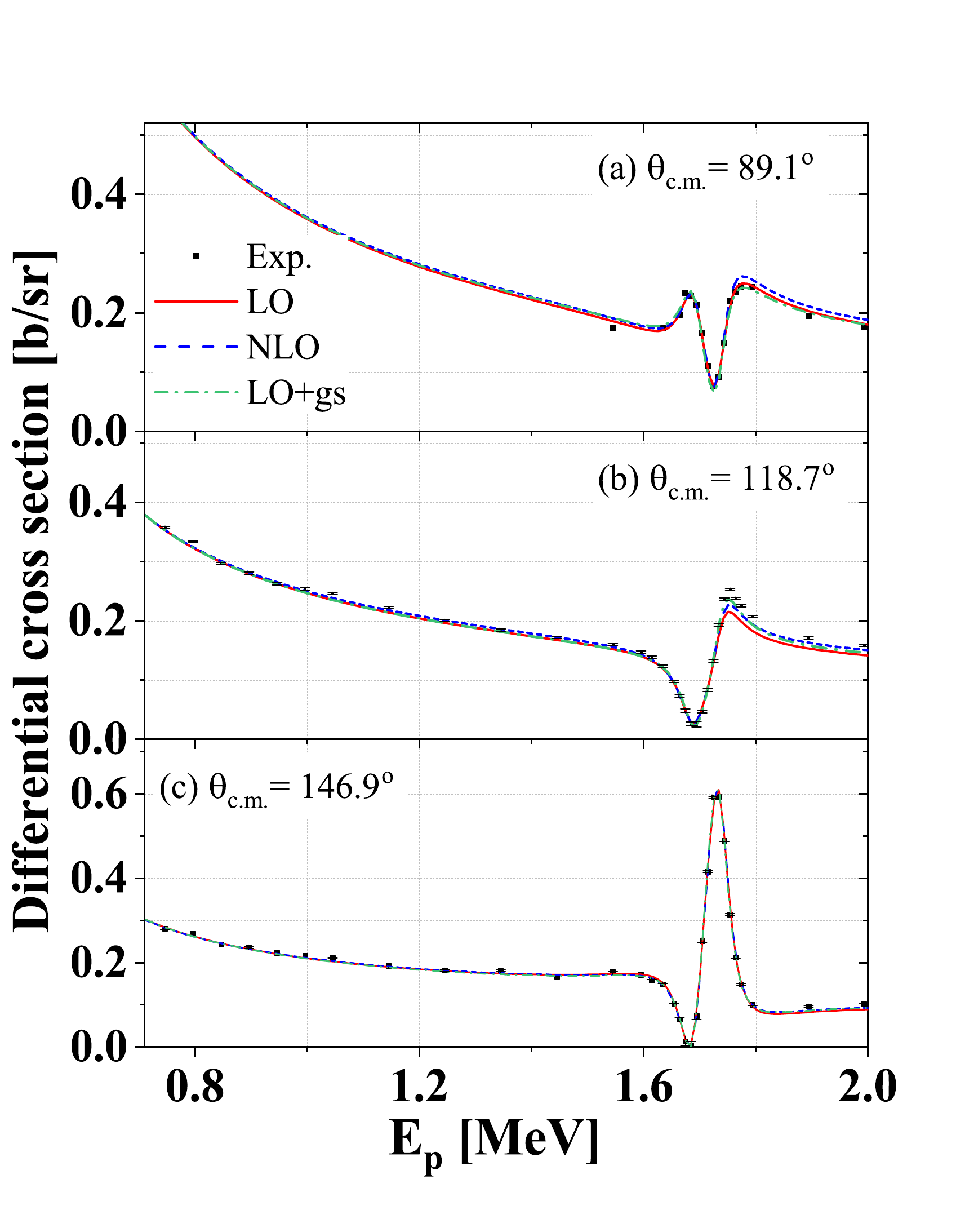}}%
\caption{\label{p12C_CShigh}
The differential cross section for elastic p-$^{12}$C scattering 
as a function of the proton energy 
$E_p= (0.7\sim 2)$ MeV
at three angles
(a) 89.1$^{\circ}$, (b) 118.7$^{\circ}$ and (c) 146.9$^{\circ}$. 
The notation is the same as in Fig.~\ref{p12C_CSlow}.
}
\end{figure}

The parameters of each set are then determined by the 
fitting to the differential cross-section data 
at three different angles,  89.1, 118.7, and 146.9 degrees~\cite{meyer1976}.
Figures~\ref{p12C_CSlow} and ~\ref{p12C_CShigh} show 
the resulting differential cross-sections
for the region $E_p \leq 0.7\ \text{MeV}$
and $E_p= (0.7\sim 2)\ \text{MeV}$,
$E_p$ is the incident proton energy.
The calculated cross sections agree very well
with the data,
which can also be seen in the obtained
 $\chi_\Lambda^2/\text{datum}=1.20$ for LO, and 1.03 for NLO.
 
%


The values of the fitted ERE parameters for the expansion around the origin 
are summarized in Table~\ref{EREpara}.
In the NLO case, compared to LO,
the added parameters,
the $P_1'$ for the $p_{3/2}$-wave and
the $P_2'$ and $Q_2'$ for the $d_{5/2}$-wave,
have large uncertainties that are much bigger than the central values.
This might be
due to
a strong correlation between the parameters of $p_{3/2}$ and $d_{5/2}$-waves,
which is not surprising since the pole positions 
at 1.686 and 1.734 MeV, respectively,
are
very close to each other. 
%

\begin{figure} [htb]
\centering
\resizebox{\columnwidth}{!}{%
\includegraphics[]{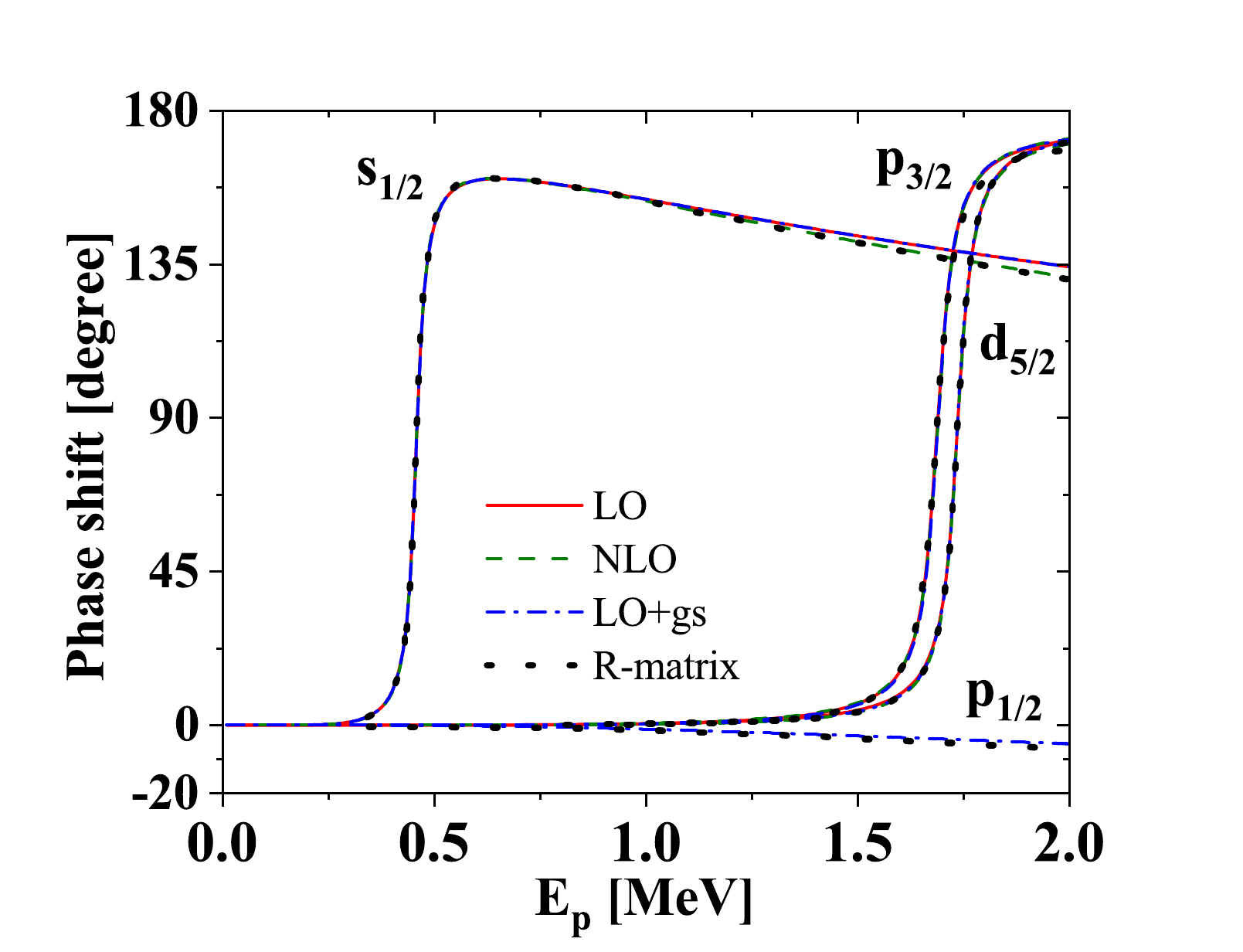}}%
\caption{\label{p12C_PS}
The phase shifts.
The red, green and blue lines are for LO, NLO and LO+gs,
respectively.
The black short-dashed lines are 
the results by using the $R$-matrix
taken from Ref.~\cite{Azu2010}.}
\end{figure}

We also considered the 
role
of the ground state on the differential cross-sections. 
Our results 
show
that including
the ground state provides a more accurate description of 
differential cross-section in high energy region, particularly around 1.7 MeV. 
Our result is in line
with the 
finding obtained from the 
$R$-matrix 
study given in Ref.~\cite{Azu2010}.
Fig.~\ref{p12C_PS} shows that the ground state
gives us a small and slowly varying
repulsive contribution.
As can be seen in Table~\ref{EREpara},
the pole position parameter for this channel
suffers from a very big uncertainty, $E_r= (-1\pm 13)$ MeV,
which is not surprising recalling that
the ground state
lies below about 1.9 MeV from the threshold.

\begin{table*}
\begin{center}
\caption{\label{EREpara}
The ERE parameters 
for the expansion around the pole positions.
}
%
\begin{tabular}{|c|r|r|r|r|}
\Xhline{3\arrayrulewidth}
      & $E_r$ (MeV) & $r_l^{'}$ (fm$^{1-2l}$)  & $P_l^{'}$ (fm$^{3-2l}$) & $Q_l^{'}$ (fm$^{5-2l}$)\\
\Xhline{3\arrayrulewidth}
\multicolumn{5}{|l|}{(a) LO \hfill{$\chi_\Lambda^2/N=1.20$}} \\ 
\hline
$s_{1/2}$ & $-0.09 \pm 0.00$ & $1.52 \pm 0.00$ &  &  \\ 
$p_{3/2}$ & $1.74 \pm 0.00$ & $-1.83 \pm 0.04$ &  & \\
$d_{5/2}$ & $1.76 \pm 0.01$ & $-0.19 \pm 0.02$ & $4.94 \pm 2.65$ &  \\ 
\hline \hline
\multicolumn{5}{|l|}{(b) NLO \hfill{$\chi_\Lambda^2/N=1.03$}} \\ 
\hline
$s_{1/2}$ & $-0.11 \pm 0.00$ & $1.46 \pm 0.01$ & $-2.18 \pm 0.11$  & \\ 
$p_{3/2}$ & $1.74 \pm 0.02$ & $-2.01 \pm 0.42$  & $25.40 \pm 55.15$  & \\ 
$d_{5/2}$ & $1.78 \pm 0.02$ & $-0.16 \pm 0.03$ & $0.12 \pm 3.83$ & 
$(3.35\pm 6.69)\times 10^{-6}$\\ 
\hline \hline
\multicolumn{5}{|l|}{(c) LO+gs \hfill{$\chi_\Lambda^2/N=1.01$}} \\ 
\hline
$s_{1/2}$ & $-0.09 \pm 0.00$ & $1.52 \pm 0.00$  & & \\ 
$p_{3/2}$ & $1.74 \pm 0.00$ & $-1.91 \pm 0.04$ &  & \\ 
$d_{5/2}$ & $1.77 \pm 0.01$ & $-0.18 \pm 0.02$ & $2.14 \pm 2.86$ & \\ 
$p_{1/2}$ & $-1.02 \pm 12.98$ & $-0.20 \pm 0.98$ &  & \\ 
\hline
\end{tabular}
\end{center}
\end{table*}


%


The phase shifts
are plotted and compared with 
the $R$-matrix analysis in Fig.~\ref{p12C_PS}. 
As can be seen in Fig.~\ref{p12C_PS},
our results are very close to 
those obtained
by $R$-matrix calculations~\cite{Azu2010}.


\clearpage
\section{Conclusions\label{Sec:Conclusions}}

The elastic p-$^{12}$C scattering 
at energies below $E_p \leq 2\ \text{MeV}$
is studied
in terms of a cluster EFT,
the pertinent degrees of freedom
of which are
the proton, the ground state of $^{12}$C
and
the three low-lying states ($s_{1/2}$, $p_{3/2}$, 
$d_{5/2}$) 
of 
$^{13}$N.
%
The resulting
scattering amplitudes of the theory are found to be consistent with the 
Coulomb-modified ERE,
and
the low-energy constants are 
represented by the ERE parameters.
At 
the leading-order, 
we have seven parameters, 2 for 
each of the $s$- and $p$-wave, and 3 for the $d$-wave.
The theory prediction turns out to be 
in a very good agreement with the
experimental data,
achieving $\chi_\Lambda^2/\text{datum}=1.20$
(see Eqs.~(\ref{chi2},\ref{Lambdai}) for 
the  definition of $\chi_\Lambda^2$).

The fitting procedure
for the ERE parameters
can be substantially simplified
by expanding the ERE around the pole positions and defining the ERE parameters accordingly,
which strongly reduces 
correlations among the parameters.
The effect of the higher-order terms 
has been studied by adding
one higher-order term 
for each partial wave,
which is denoted as NLO
and  scores $\chi_\Lambda^2/\text{datum}=1.03$.
%
To estimate 
the role of the
ground state of ${}^{13}$N 
that lies below the threshold,
we have also considered the cases where
the ground state is promoted to
an explicit degree of freedom.
The resulting ``LO+gs" theory 
results in $\chi_\Lambda^2/\text{datum}=1.01$.
These improvements of NLO and LO+gs
are, however, accompanied
by large uncertainties in the
additionally introduced
ERE parameters (see Table~1).
It shows that
the experimental data 
considered in this work
with $E_p \le 2$ MeV
are well described by the LO,
and 
the contributions from the higher order
terms and the sub-threshold ground state
are not essential. 
%
The resulting phase shifts are in 
an excellent agreement with the
$R$-matrix analysis~\cite{Azu2010}.

The high momentum scale of a
low-energy
EFT is set by the
lowest-energy state that 
is not taken explicitly,
the $2^+$ state of $^{12}$C with $E_x=4.44\ \text{MeV}$.
This 
corresponds 
to rather a large expansion parameter $k_{lo}/k_{hi}\,=\,(0.3 \sim 0.6)$.
The main mechanism that makes our 
approach
successful 
despite this rather large ratio
might be traced to the relevance of the ERE at low energies.
An immediate extension of this work would be
${}^{12}\mathrm{C}(p,\gamma)^{13}\mathrm{N}$.

%


\section*{Acknowledgments}

This work was supported in part by the Korean government 
Ministry of Science and ICT (MSIT)
through the National Research Foundation 
(2020R1A2C110284)
and in part by the Institute for Basic Science (IBS-R031-D1).
The work of Y.-H.S. was 
supported by the Rare Isotope Science Project (RISP) of Institute for Basic Science (IBS) funded by 
the MSIT through the NRF
(2013M7A1A1075764) and 
by the National Supercomputing Center with supercomputing resources including technical support
(KSC-2020-CRE-0027).
The work of E.J.I. was prepared in part
by LLNL
under Contract DE-AC52-07NA27344.


%

\end{document}